\documentclass[manuscript,screen,oneside]{acmart}

\AtBeginDocument{%
  }

\renewcommand\footnotetextcopyrightpermission[1]{}
\makeatletter
\@twosidefalse
\makeatother

\usepackage{graphicx}       
\usepackage{acronym}    
\usepackage{array}
\usepackage{longtable}
\usepackage{booktabs}
\usepackage{multirow}
\usepackage{tabularx}
\usepackage{appendix}
\usepackage{hyperref}       
\usepackage{xcolor}
\usepackage{todonotes}

\usepackage{verbatim}

\begin{document}

\newcommand{\alessio}[1]{\todo[color=blue!20, inline]{\textbf{Alessio:} #1}}
\newcommand{\tom}[1]{\todo[color=orange!20, inline]{\textbf{Tom:} #1}}

\makeatletter
\def\@fnsymbol#1{\ensuremath{\ifcase#1\or *\or \dagger\or \ddagger\or 
   \mathsection\or \mathparagraph\or \|\or Target\or \dagger\dagger 
   \or \ddagger\ddagger \else\@ctrerr\fi}}
\makeatother

\title{Assessing High-Risk AI Systems under the EU AI Act: From Legal Requirements to Technical Verification}

\author{Alessio Buscemi$^{1}$}
\affiliation{%
  \institution{Luxembourg Institute of Science and Technology (LIST)}
  \country{Luxembourg}
}

\author{Tom Deckenbrunnen$^{1}$}
\affiliation{%
  \institution{LIST, University of Luxembourg}
  \country{Luxembourg}
}

\author{Fahria Kabir$^{2}$}
\author{Kateryna Mishchenko$^{2}$}
\author{Nishat Mowla$^{2}$}

\affiliation{%
  \institution{Research Institutes of Sweden (RISE)}
  \country{Sweden}
}

\email{1{name}.{surname}@list.lu}
\email{2{name}.{surname}@ri.se}


\acrodef{AI}{Artificial Intelligence}
\acrodef{FCA}{Financial Conduct Authority} 
\acrodef{AI Act}{Artificial Intelligence Act}
\acrodef{GDPR}{General Data Protection Regulation}
\acrodef{CRA}{Cyber Resilience Act} 
\acrodef{SME}{Small-medium enterprise}
\acrodef{GPAI}{General Purpose AI}
\acrodef{DGA}{Data Governance Act}
\acrodef{DSA}{Digital Services Act}
\acrodef{DMA}{Digital Markets Act} 
\acrodef{AIRS}{AI Regulatory Sandbox}
\acrodef{AITS}{AI Technical Sandbox}
\acrodef{EUSAIR}{EU Regulatory Sandboxes for AI}
\acrodef{LIST}{Luxembourg Institute of Science and Technology} 
\acrodef{CNPD}{National Data Protection Commission}
\acrodef{EDIH}{European Digital Innovation Hub} 
\acrodef{TEF}{testing and Experimentation Facilities}
\acrodef{EuroHPC}{European High-Performance Computing}
\acrodef{EDIC}{European Digital Infrastructure Consortia}
\acrodef{AIoD}{AI-on-Demand Platform}
\acrodef{AI RMF}{AI Risk Management Framework}
\acrodef{CA}{Competent Authority}
\acrodef{NIST}{National Institute of Standards and Technology}
\acrodef{RMF}{Risk Management Framework}
\acrodef{CEN}{European Committee for Standardization}
\acrodef{CENELEC}{European Committee for Electrotechnical Standardization}
\acrodef{JTC}{Joint Technical Committee}
\acrodef{RAG}{Retrieval Augmented Generation}
\acrodef{NLP}{Natural Language Processing}
\acrodef{HPC}{High Performance Computing}
\acrodef{DSL}{Domain-Specific Language}
\acrodef{RISE}{Research Institutes of Sweden AB}
\acrodef{HLEG}{High-Level Expert Group on Artificial Intelligence}
\acrodef{DORA}{Digital Operational Resilience Act}

\begin{abstract}
The implementation of the AI Act requires practical mechanisms to verify compliance with legal obligations, yet concrete and operational mappings from high-level requirements to verifiable assessment activities remain limited, contributing to uneven readiness across Member States. This paper presents a structured mapping that translates high-level AI Act requirements into concrete, implementable verification activities applicable across the AI lifecycle.
The mapping is derived through a systematic process in which legal requirements are decomposed into operational sub-requirements and grounded in authoritative standards and recognised practices. From this basis, verification activities are identified and characterised along two dimensions: the type of verification performed and the lifecycle target to which it applies.
By making explicit the link between regulatory intent and technical and organisational assurance practices, the proposed mapping reduces interpretive uncertainty and provides a reusable reference for consistent, technology-agnostic compliance verification under the AI Act.
\end{abstract}



\maketitle
\section{Introduction}
\label{sec:introduction}

Progress in AI has created substantial opportunities while also introducing significant technical, organisational, and regulatory challenges, particularly as AI systems are increasingly deployed in critical domains where failures may have serious societal consequences \cite{bengio2024managing, wang2022artificial}. In response, regulatory efforts worldwide are intensifying to promote trustworthiness in AI development and deployment \cite{huang2024, smuha2021race, novelli2024taking}. Within this context, the EU \ac{AI Act} \cite{eu_ai_act_2024} represents one of the most comprehensive regulatory responses to date, establishing a risk-based framework that mandates ex ante verification and ongoing oversight for high-risk AI systems.
Despite this progress, a persistent gap remains between high-level legal requirements and the concrete verification activities needed to demonstrate compliance in practice. This gap reflects the absence of explicit and shared mappings between regulatory obligations and implementable assurance practices, resulting in fragmented compliance efforts across stakeholders and Member States and limiting comparability and regulatory learning. In practice, organisations often struggle to translate abstract legal obligations into verifiable controls, tests, and documentation artefacts that can be applied consistently across the AI lifecycle.

These challenges are compounded by three interrelated forms of uncertainty. Interpretive uncertainty concerns the contextual meaning and scope of legal obligations; operational uncertainty relates to how such obligations can be instantiated across heterogeneous development and deployment practices; and procedural uncertainty arises from evolving standards, guidance documents, and the gradual designation of notified bodies \cite{deloitte2024ai, dlapiper2025pause, draghi2024competitiveness, arnal2024implementation, lewis2025regulatorylearning, regulatinguncertainty2025}. These uncertainties are further amplified by the emergent behaviour of modern generative AI systems, whose properties arise from complex interactions rather than explicit programming \cite{holtzmanGenerativeModelsComplex2025, koltLessonsComplexSystems2025a}. As a horizontal regulation relying on harmonised standards, post-market monitoring, and iterative guidance, the AI Act embodies a model of regulatory learning in which compliance expectations are progressively refined through implementation experience and verification outcomes \cite{zakiConceptualisingOrganisationalPolicy2025, lewis2025regulatorylearning, schrepelAdaptiveRegulation2025, mowla2024ai}. Such learning depends on the availability of structured, comparable, and traceable verification evidence, which remains fragmented across actors, disciplines, and use cases.

This paper addresses the need for an explicit and operational mapping between AI Act requirements and concrete verification activities that can be executed, documented, and compared across the AI lifecycle. The objective is to operationalise compliance by making the link between regulatory intent and technical and organisational assurance practices explicit, reproducible, and auditable. The focus is on high-risk AI systems as defined in Article~6 and Annex~III, which concern application domains where failures may adversely affect health, safety, or fundamental rights. Compliance must be demonstrated through conformity verification procedures under Articles~43 and~44, either via internal control procedures (Annex~VI) or third-party assessment by notified bodies (Annex~VII), both requiring verifiable evidence spanning design, development, deployment, and post-market monitoring.

Translating these obligations into concrete assessment activities remains challenging. Many provisions of the AI Act are inherently context-dependent, requiring case-specific interpretation grounded in the system’s intended use, operational environment, and evolving risk profile. Requirements related to data governance, robustness, accuracy etc. encompass multiple dimensions, each necessitating distinct validation and testing methodologies, while the lifecycle-oriented nature of the Act entails continuous monitoring and post-deployment evaluation rather than reliance on point-in-time certification alone. Existing initiatives provide partial foundations for AI verification, but none offers a complete and operational mapping between legal requirements and verification activities. The NIST AI Risk Management Framework \cite{NIST_AI_RMF_2025} provides high-level guidance but is not aligned with EU-specific legal obligations, while ISO/IEC standards such as ISO/IEC~42001 \cite{iso_iec_42001_2023} establish process-oriented management requirements without specifying AI Act–specific verification procedures. European standardisation efforts within CEN--CENELEC JTC~21 \cite{CEN-CENELEC_JTC21_2025} aim to develop harmonised standards granting presumption of conformity under Article~40, but this work remains ongoing.

Against this background, the central contribution of this paper is a structured and operational mapping that translates high-level AI Act requirements into concrete verification activities applicable across the AI lifecycle. The mapping is constructed through a systematic decomposition of legal obligations into operational sub-requirements and is grounded in authoritative standards and recognised engineering and governance practices. To ensure consistency and comparability, verification activities are characterised along two dimensions capturing both verification type and lifecycle target. The applicability of the mapping is illustrated through a real-world case study involving a high-risk AI system in the automotive domain.
Within this scope, the paper addresses three guiding questions that structure the construction of the proposed mapping: (i) how high-level legal obligations can be decomposed into concrete verification activities; (ii) which verification dimensions are required to ensure comparability and traceability across heterogeneous actors; and (iii) how a shared verification structure can support communication and regulatory learning across use cases and sectors. While related efforts exist (e.g. \cite{hernandez2025open}), the proposed mapping provides a higher level of granularity and operationalisability. It is not intended to be prescriptive or exhaustive, but to serve as a reusable reference that can evolve alongside regulatory guidance, standardisation efforts, and advances in evaluation techniques, thereby supporting more consistent conformity verification and evidence-based regulatory learning over time.

\section{Methodology}
\label{sec:methodology}

\begin{figure}[!htb]
    \centering
    \includegraphics[width=1\linewidth]{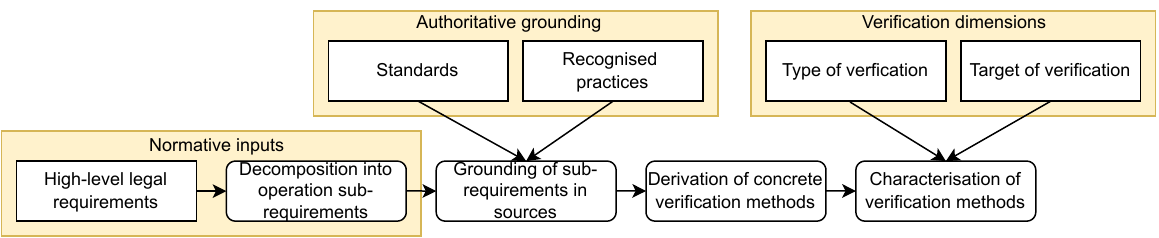}
    \caption{Overview of the methodology.}
    \label{fig:framework}
    \vspace{-6mm}
\end{figure}

This section describes the methodology used to construct a structured mapping between high-level AI Act requirements and concrete, implementable verification activities.
Figure~\ref{fig:framework} provides an overview of the resulting compliance verification structure and the main methodological building blocks. It illustrates how high-level legal requirements are progressively decomposed, grounded, and translated into concrete verification activities, which are then situated within a structured verification space defined along two core dimensions.

\subsection{Normative inputs and requirement structuring}
\label{sub:requirements}

The starting point of the methodology consists in identifying and structuring a set of high-level requirements derived from the obligations established by the AI Act. Given the heterogeneity and dispersion of these obligations across the regulation, an explicit structuring step is required to organise them into a coherent form suitable for systematic verification.
For this purpose, we use the seven principles of Trustworthy AI defined by the European Commission’s High-Level Expert Group in the Ethics Guidelines for Trustworthy AI~\cite{eu_trustworthy_ai_2019} as an organising abstraction. Although these principles predate the AI Act and are not legally binding, they have strongly influenced the development of European AI governance and are reflected throughout the regulation, both as explicit obligations (e.g. transparency, human oversight) and as distributed requirements embedded in provisions on governance, documentation, and system performance. Their role in the methodology is therefore structural: they serve as a stable taxonomy for grouping and analysing legally binding requirements, not as an independent source of obligations.

To account for the more prescriptive and organisational character of the AI Act, this abstraction is complemented with four additional requirement categories corresponding to obligations that are central to conformity assessment for high-risk systems: quality management, risk management, technical documentation, and record keeping. These categories capture lifecycle governance and accountability mechanisms that are explicitly mandated by the regulation but are only partially represented in principle-based formulations. Together, the resulting eleven high-level requirements provide a structured normative representation of the obligations imposed by the AI Act for high-risk systems. Each requirement corresponds to a coherent cluster of legally binding provisions, organised in a form suitable for systematic verification and without introducing additional normative content beyond the regulation itself.

Each high-level requirement is subsequently decomposed into more granular operational components. These components correspond either to obligations explicitly stated in the AI Act or to requirements that can be reasonably derived from its provisions in order to enable verification. This decomposition does not aim to provide an exhaustive or definitive interpretation of the law, but to articulate traceable operational interpretations that can be subjected to assessment.
The methodology explicitly allows for alternative decompositions depending on context, sector, and risk profile, and is designed to accommodate such variability rather than eliminate it.

\subsection{Authoritative grounding}
\label{sub:grounding}

To ensure both normative alignment and technical soundness, the operational interpretation of requirements is grounded in authoritative sources that specify or exemplify how high-level obligations are implemented and assessed in practice. These sources are predominantly non-binding, complemented where appropriate by binding instruments whose established operational mechanisms are widely reused in compliance and assurance activities.
In this respect, the GDPR~\cite{GDPR2016} is included not as a source of normative extension of the AI Act, but because it provides mature and widely operationalised mechanisms—such as risk-based assessment, documentation practices, and accountability procedures—that directly inform how several AI Act obligations are interpreted and verified in practice.
The AI Act explicitly operates in complementarity with existing data protection law, and high-risk AI systems in the European context frequently involve personal data processing throughout development, deployment, and monitoring. As a result, verification activities related to data governance, transparency and information provision, record keeping, and affected-person rights often rely on GDPR-derived concepts, artefacts, and procedures, such as information duties, accountability documentation, and impact-assessment practices. For this reason, the GDPR is treated here as an authoritative operational reference within the grounding layer.

Beyond this exception, operational grounding relies on non-binding but authoritative sources that provide reusable mechanisms for implementing and assessing regulatory obligations in practice. These include the EU General Purpose AI Code of Practice~\cite{gpai-code-practice-3rd}, ISACA’s Advanced Audit in AI Official Review Manual~\cite{isaca2025aaia}, UNESCO’s Recommendation on the Ethics of AI~\cite{unesco2021ethics}, the Web Content Accessibility Guidelines (WCAG)~2.2~\cite{wcag22}, and a broad range of ISO and IEC standards relevant to AI governance, risk management, quality management, cybersecurity, and data quality (e.g.~\cite{ISO_IEC_22989_2022, ISO23894, ISO_IEC_TR_24027_2021, ISO_IEC_24029_2_2023, ISO_IEC_24970_DIS_2025, ISO_IEC_27001_2022, ISO_IEC_27005_2022, ISO_IEC_27035_1_2023, ISO9001, ISO_31000_2018, ISO_IEC_31010_2019, iso_iec_42001_2023, ISO_IEC_5259_1_2024, ISO_8000_8_2015}). These sources were selected because they translate high-level principles into implementable governance processes, controls, and audit practices, thereby bridging normative requirements and operational verification.

Where regulatory, guidance, and standardisation sources did not sufficiently specify technical verification procedures, the analysis was extended to the scientific literature. Highly cited and influential works in AI governance, robustness testing, risk management, and human–system interaction were considered (e.g.~\cite{Brundage2020, Basiri2016, Floridi2018, mundhenk2020, sheridan1992telerobotics, schuler1993participatory, CoE_AI_2020}). Selection criteria combined citation impact, venue relevance, and explicit connection to measurable or testable aspects of AI trustworthiness.
This layered grounding process ensures that each operational requirement and associated verification activity can be traced back to authoritative legal, standardisation, or scientific sources.

\subsection{Verification dimensions}
\label{sub:dimensions}

\begin{figure}[!htb]
    \centering
    \includegraphics[width=0.4\linewidth]{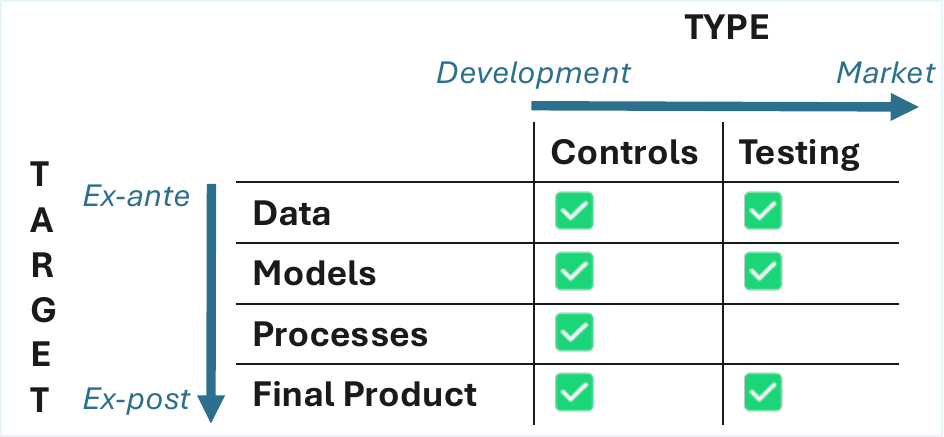}
    \caption{Verification space defined by verification type and verification target.}
    \label{fig:dimensions}
    \vspace{-3mm}
\end{figure}

Concrete verification activities are organised within a structured verification space defined by two orthogonal dimensions: the type of verification and the target of verification within the AI lifecycle. Together, these dimensions provide a common operational language for identifying, organising, and comparing verification activities across heterogeneous systems and stakeholders.
The first dimension concerns the \textit{type of verification}. Two broad categories are distinguished: controls and testing. Controls refer to process-based assurance mechanisms, including governance structures, documentation practices, quality management procedures, and organisational safeguards that ensure responsible system development and operation. Testing refers to empirical evaluations that assess system behaviour, performance, or properties under specified conditions. Controls and testing are complementary rather than hierarchical: controls establish accountability and traceability, while testing provides empirical evidence of compliance. Both may be applied at any stage of the AI lifecycle.
The second dimension concerns the \textit{target of verification}, corresponding to the lifecycle component or artefact being evaluated. Numerous software and AI lifecycle models provide fine-grained decompositions of development and deployment phases (e.g. \cite{de2022artificial, wang2021artificial}). However, such models vary significantly across standards, engineering practices, and organisational contexts, and their level of detail often presupposes technical expertise. For the purpose of compliance verification, which involves regulators, auditors, legal experts, risk managers, and developers alike, a higher level of abstraction is required to support immediate interpretability and shared understanding.

Accordingly, the methodology adopts four high-level verification targets that act as stable and intuitive aggregation buckets rather than exhaustive lifecycle stages. The \textit{data} target addresses issues related to data acquisition, composition, representativeness, and quality. The \textit{model} target concerns the AI system’s architecture, training procedures, explainability, robustness, and performance characteristics. The \textit{processes} target encompasses organisational and procedural elements such as risk management, quality assurance, documentation, traceability, and lifecycle governance. The \textit{final product} target focuses on the behaviour of the deployed system, including its interaction with users, integration into sociotechnical contexts, and observable outputs and impacts.
This abstraction is intentional: it sacrifices fine-grained lifecycle specificity in favour of clarity, stability, and cross-disciplinary usability. Each target is sufficiently broad to accommodate different development methodologies and technical implementations, while remaining concrete enough to support systematic verification and evidence collection.

While analytically distinct, these targets may overlap in practice. Certain verification activities, such as model monitoring or data versioning, span multiple targets simultaneously. The methodology treats these boundaries as pragmatic rather than absolute, allowing verification activities to address multiple concerns where appropriate.
Within this verification space, each high-level requirement is associated with a set of control and testing activities consistent with its legal intent and operational scope. This structure enables evaluators to systematically identify relevant verification activities, organise evidence coherently, and ensure coverage across governance and empirical dimensions. By making explicit how regulatory requirements translate into concrete verification actions, the methodology supports comparability, traceability, and communication across heterogeneous stakeholders and use cases.

\section{Mapping}
\label{sec:mapping}

In this section, we present the mapping between the high-level requirements of the AI Act defined in Section~\ref{sub:requirements} and the mechanisms that support their implementation and verification. Each subsection (\textit{R1–R11}) introduces a brief explanation of the corresponding requirement, followed by the actual mapping presented in the form of a table. These tables identify the relevant legal provisions, associated authoritative references, and the concrete methods through which compliance can be assessed or demonstrated.
To ensure conceptual clarity and consistency across requirements, each method is classified along the two analytical dimensions described in Section~\ref{sub:dimensions}. \textit{Type} distinguishes between \textit{C (Controls)} and \textit{T (Tests)}, while \textit{Target} specifies the primary layer of the AI lifecycle to which the method applies: \textit{D (Data)}, \textit{M (Models)}, \textit{P (Processes)}, and \textit{FP (Final Product)}.

\subsection{R1: Human Agency and Oversight}

Article~14 of the AI Act establishes the principle of human agency and oversight as a cornerstone of trustworthy AI. It requires that AI systems be designed and implemented so that they operate under meaningful human control and that ultimate responsibility remains with human operators. In practice, this means that individuals supervising the system must understand its capabilities and limitations, be able to intervene when necessary, and override its decisions to prevent or mitigate risks.

{\footnotesize
\begin{longtable}{|p{2.1cm}|p{2cm}|p{8cm}|p{0.8cm}|p{0.8cm}|}
\hline
\textbf{Requirement} & \textbf{References} & \textbf{Methods} & \textbf{Type} & \textbf{Target} \\
\hline
R1.1: Human in the loop & Sheridan et al. \cite{sheridan1992telerobotics} & M1.1: Manual override — Confirm that a human operator can take control of the AI system at any time during operation. & C & P, FP \\ \cline{2-5}
 & Sheridan et al. \cite{sheridan1992telerobotics} & M1.2: Supervisory control — Assess whether human operators are able to supervise the AI system and intervene when necessary. & C & P, FP \\ \cline{2-5}
 & ISO/IEC 42001:2023 (Ann.A.3.2) & M1.3: Strategic governance — Ensure that organisational oversight mechanisms for AI deployment are defined and documented. & C & P \\ \hline
R1.2: User Autonomy & GDPR (Art.7); Andreotta et al. \cite{andreotta2022ai} & M1.4: Informed consent — Check that users are provided with clear information about AI usage and are able to give or withhold consent. & C & FP \\ \cline{2-5}
 & ISACA manual (\S~3.2) & M1.5: User preferences — Review whether the system allows adaptation of its behaviour based on user-set preferences. & C & FP \\ \cline{2-5}
 & ISACA manual (\S~3.2) & M1.6: Opt-in/out — Confirm that users can enable or disable AI functionalities according to their choices. & C & FP \\ \hline
R1.3: Oversight Mechanisms & ISACA manual (\S~2.11.1) & M1.7: Audit trails — Ensure that system actions and decisions are recorded in tamper-evident logs for traceability. & C & P \\
\hline
\end{longtable}
\vspace{-3mm}
}

\subsection{R2: Technical Robustness and Safety}

Technical robustness and safety under the AI Act (Article 15) concern the system's ability to operate reliably under normal and adverse conditions, to resist manipulation or degradation, and to recover safely from failures. These requirements ensure that AI systems are resilient to risks arising from data drift, adversarial attacks, or component malfunctions, and that their performance remains consistent throughout the lifecycle. They also mandate mechanisms for fallback and uncertainty management, so that any deviation from intended operation can be detected, controlled, and mitigated in a way that preserves both functionality and safety.

{\footnotesize
\begin{longtable}{|p{2.1cm}|p{2cm}|p{8cm}|p{0.8cm}|p{0.8cm}|}
\hline
\textbf{Requirement} & \textbf{References} & \textbf{Methods} & \textbf{Type} & \textbf{Target} \\
\hline
R2.1: Resilience \& Reliability & ISACA manual (\S~1.4.4) & M2.1: Stress testing — Assess whether the system has been exposed to extreme loads or edge cases to identify potential failure points. & T & P \\ \cline{2-5}
 & ISACA manual (\S~2.8.6) & M2.2: Drift detection — Confirm that mechanisms are in place to monitor data distribution changes and evaluate behaviour under such shifts. & T & D, P \\ \cline{2-5}
 & Basiri et al.~\cite{Basiri2016} & M2.3: Chaos engineering — Determine whether resilience has been tested through controlled fault injection in production-like environments. & T & P \\ \cline{2-5}
 & ISACA manual (\S~2.9.4) & M2.4: Fail-over architecture — Ensure redundant components exist to maintain functionality in case of failure. & C & P, FP \\ \hline
R2.2: Robustness to Attacks & Brundage et al.~\cite{Brundage2020}; ISACA manual (\S~2.13.3) & M2.5: Red teaming — Examine whether adversarial testing is conducted to uncover vulnerabilities. & T & P, FP \\ \hline
R2.3: Fallback \& Fail-safe & HLEG Guidelines & M2.6: Human fallback — Check that the system includes a mechanism to transfer control to a human in the event of failure. & C & P, FP \\ \hline
R2.4: Accuracy \& Uncertainty & ISACA manual (\S~1.4.3) & M2.7: Model calibration — Review whether confidence scores are appropriately aligned with model performance. & T & M \\
\hline
\end{longtable}
\vspace{-3mm}
}

\subsection{R3: Privacy and Data Governance}

Article~10 of the AI Act establishes specific obligations concerning data governance and data quality for high-risk AI systems, requiring that datasets be relevant, representative, free of errors, and complete to the extent possible. These obligations are AI-specific and apply irrespective of whether personal data is involved. 
Where high-risk AI systems rely on personal data, the operational interpretation and verification of Article~10 requirements necessarily draw on established data protection mechanisms defined under the GDPR, in particular Articles~5, 25, and~32, which address purpose limitation, data protection by design, and security of processing. In such cases, GDPR concepts and safeguards provide concrete, widely operationalised mechanisms through which certain data governance obligations of the AI Act can be implemented and assessed in practice.
Accordingly, compliance with Article~10 typically involves a combination of organisational and technical measures, such as encryption, pseudonymisation, access controls, and accountability procedures, which are already well established in data protection practice. Effective data governance thus supports compliance with AI Act requirements while also contributing to the technical robustness and trustworthiness of AI systems across their lifecycle.

{\footnotesize
\begin{longtable}{|p{2.1cm}|p{2cm}|p{8cm}|p{0.8cm}|p{0.8cm}|}
\hline
\textbf{Requirement} & \textbf{References} & \textbf{Methods} & \textbf{Type} & \textbf{Target} \\
\hline
R3.1: Data Minimisation & GDPR (Art.~5.1.b–c) & M3.1: Purpose limitation — Verify that only data strictly necessary for a defined processing purpose is collected. & C & D, P \\ \cline{2-5}
 & GDPR (Art.~5.1.b) & M3.2: Scope restriction — Assess whether processing is confined to explicitly defined contexts. & C & D, P \\ \hline
R3.2: Privacy-Enhancing & GDPR (Art.~25.1–2; Art.~32.1.a) & M3.3: Pseudonymisation — Check whether mechanisms are in place to pseudonymise personal data. & C & D \\ \hline
R3.3: Access \& Security & GDPR (Art.~32.1.b–d) & M3.4: Role-based access control — Ensure access rights are granted based on organisational roles. & C & P \\ \cline{2-5}
 & GDPR (Art.~32.1.a); ISACA manual (\S~2.7.3) & M3.5: Encryption — Check that data is encrypted in storage and transmission. & T & D, P \\ \hline
R3.4: Data Provenance & ISACA manual (\S~2.7.5) & M3.6: Lineage tracking — Confirm that data transformations are logged to ensure traceability. & C & D, P \\ \cline{2-5}
 & ISACA manual (\S~3.1.2) & M3.7: Metadata versioning — Verify dataset versions and schema changes are tracked. & C & D, P \\ \hline
R3.5: Consent Management & GDPR (Art.~7.1–2) & M3.8: Consent tracking — Inspect whether user consents are explicitly obtained, logged, and tied to specific purposes. & C & P, FP \\ \cline{2-5}
 & GDPR (Art.~7.3) & M3.9: Granular revocation — Confirm that users can revoke permissions without affecting unrelated consents. & C & FP, P \\
\hline
\end{longtable}
}

\subsection{R4: Transparency}

Transparency is one of the most fundamental requirements for trustworthy and accountable AI. Under the AI Act, specifically in Art. 13, AI systems must be designed and documented in a way that allows users, regulators, and auditors to understand their capabilities, limitations, and decision-making logic. Transparency encompasses both technical and communicative aspects: it includes model explainability, data and process traceability, disclosure of AI usage to end users, and interpretability of outputs in accessible language. Together, these mechanisms ensure that AI operations remain observable and interpretable across all stages of their lifecycle, thereby fostering accountability and enabling meaningful human oversight.

{\footnotesize
\begin{longtable}{|p{2.1cm}|p{2cm}|p{8cm}|p{0.8cm}|p{0.8cm}|}
\hline
\textbf{Requirement} & \textbf{References} & \textbf{Methods} & \textbf{Type} & \textbf{Target} \\
\hline
\multirow{2}{=}{R4.1: Explainability} 
 & ISACA manual (\S~3.2) 
 & M4.1: Local interpretability — Verify that the system provides human-understandable explanations of model outputs using recognised local explanation techniques (e.g., feature attribution or surrogate models). 
 & T & D, M, FP \\ \cline{2-5}
 & Mundhenk et al.~\cite{mundhenk2020} 
 & M4.2: Saliency maps — Check if visualisation tools highlight relevant input regions. 
 & T & D, M, FP \\ \hline

\multirow{2}{=}{R4.2: Model \& Data Traceability} 
 & GPAI CoP (Sec. Transparency) 
 & M4.3: Model cards — Ensure that standardised documentation is available and maintained. 
 & C & M, P \\ \cline{2-5}
 & GPAI CoP (Sec. Transparency) 
 & M4.4: Datasheets — Assess whether datasheets describing dataset origin and composition are accessible. 
 & C & D, P \\ \hline

\multirow{3}{=}{R4.3: Disclosure of AI Use} 
 & ISO/IEC 22989 (\S 5.15) 
 & M4.5: AI identity notices — Verify that users are clearly informed when interacting with an AI system. 
 & C & FP \\ \cline{2-5}
 & GPAI CoP; ISACA manual (\S~3.2) 
 & M4.6: Disclaimers — Check whether limitations and assumptions are transparently disclosed. 
 & C & FP \\ \cline{2-5}
 & ISO/IEC 23894 (\S 6.1) 
 & M4.7: Risk communication — Assess whether the system communicates known risks and uncertainties. 
 & C & FP \\ \hline

R4.4: Interpretability for End Users 
 & ISO/IEC 22989 (\S 5.15) 
 & M4.8: Understandable language — Confirm that technical terms are explained using plain language. 
 & C & FP \\ \hline
\end{longtable}
\vspace{-3mm}
}

\subsection{R5: Diversity, Non-discrimination and Fairness}

Diversity, non-discrimination, and fairness are central to ensuring that AI systems respect fundamental rights and avoid reinforcing structural biases. Articles~10, and~27 of the AI Act require that systems be designed, trained, and deployed in a way that prevents discriminatory outcomes and ensures inclusive access. This includes assessing data representativeness, applying fairness metrics across protected groups, and incorporating participatory design practices that reflect societal diversity. Fairness thus spans both technical and organisational domains, linking data quality, model evaluation, and stakeholder engagement to the ethical and legal accountability of AI systems.

{\footnotesize
\begin{longtable}{|p{2.1cm}|p{2cm}|p{8cm}|p{0.8cm}|p{0.8cm}|}
\hline
\textbf{Requirement} & \textbf{References} & \textbf{Methods} & \textbf{Type} & \textbf{Target} \\
\hline
R5.1: Fairness in Data & ISACA manual (\S~3.2) & M5.1: Bias detection — Evaluate whether datasets are examined for representation gaps. & T & D, P \\ \cline{2-5}
 & ISACA manual (\S~3.2) & M5.2: Sampling strategies — Confirm that sampling ensures balanced representation. & C & D, P \\ \hline
R5.2: Fairness in Algorithms & ISACA manual (\S~3.2) & M5.3: Protected-class fairness metrics — Verify that fairness metrics are applied across protected groups. & T & D, M, P \\ \hline
R5.3: Inclusive Design & WCAG 2.2 & M5.4: Accessibility — Check that the system supports assistive technologies. & C & FP \\ \hline
R5.4: Stakeholder Engagement & Schuler et al. \cite{schuler1993participatory} & M5.5: Participatory design — Assess whether affected users and communities are involved in design and testing. & C & FP, P \\
\hline
\end{longtable}
\vspace{-3mm}
}

\subsection{R6: Societal and Environmental Well-being}

Societal and environmental well-being extend the notion of trustworthy AI beyond individual rights, addressing collective and long-term impacts. Articles~27 and~40 of the AI Act require that systems be aligned with broader ethical and sustainability goals, including environmental protection, energy efficiency, and respect for human dignity. This involves monitoring resource consumption, integrating human rights principles into design, and ensuring that ethical oversight bodies are established to guide responsible innovation. These requirements embed sustainability and societal benefit as integral dimensions of AI governance.

{\footnotesize
\begin{longtable}{|p{2.1cm}|p{2cm}|p{8cm}|p{0.8cm}|p{0.8cm}|}
\hline
\textbf{Requirement} & \textbf{References} & \textbf{Methods} & \textbf{Type} & \textbf{Target} \\
\hline
R6.1: Environmental Impact & UNESCO (Rec.~84); ISACA manual (\S~1.17.7) & M6.1: Energy-use tracking — Verify that the system monitors computational energy consumption and carbon footprint. & T & P, FP \\ \hline
R6.2: Ethical Alignment & Leslie et al. \cite{CoE_AI_2020} & M6.2: Rights-based system — Assess whether system design aligns with fundamental rights. & C & P \\ \cline{2-5}
 & UNESCO (Rec.~58) & M6.3: Independent ethics committee — Verify that an independent ethics committee or officer is appointed. & C & P \\
\hline
\end{longtable}
\vspace{-3mm}
}

\subsection{R7: Accountability}

Accountability ensures that responsibility for AI-related outcomes is clearly defined and traceable across the system's lifecycle. Articles~12 and~17 of the AI Act establish the duty to maintain documentation, assign responsibilities, and ensure that auditability and redress mechanisms are in place. Accountability mechanisms link technical and organisational controls to transparency, enabling effective oversight and redress when harm occurs. They require clear responsibility allocation, secure logging, version-controlled documentation, and channels for incident reporting, thereby operationalising the principle that accountability cannot be delegated to the machine.

{\footnotesize
\begin{longtable}{|p{2.1cm}|p{2cm}|p{8cm}|p{0.8cm}|p{0.8cm}|}
\hline
\textbf{Requirement} & \textbf{References} & \textbf{Methods} & \textbf{Type} & \textbf{Target} \\
\hline
R7.1: Responsibility Assignment & ISACA manual (\S~3.2) & M7.1: RASCI charts — Confirm that responsibility matrices define roles for each lifecycle activity. & C & P \\ \hline
R7.2: Auditability & ISO/IEC 42001:2023 (Ann.B.6.2.8); ISACA manual (\S~2.11.1) & M7.2: Logging — Verify that detailed, tamper-evident logs of system operations are maintained and reviewed. & C & P \\ \cline{2-5}
 & ISO/IEC 42001:2023 (Ann.B.8.3) & M7.3: External reporting interfaces — Ensure that secure portals are available for third-party auditors. & C & P \\ \hline
R7.3: Incident Reporting & Floridi et al.~\cite{Floridi2018} & M7.4: Redress systems — Check whether users can report issues or contest AI-driven outcomes. & C & FP, P \\ \hline
R7.4: Documentation Integrity & ISO/IEC 42001:2023 (Ann.B.6.2.5); ISACA manual (\S~2.11.1) & M7.5: Version-controlled records — Assess whether documentation is versioned and traceable over time. & C & P \\
\hline
\end{longtable}
\vspace{-3mm}
}

\subsection{R8: Quality Management}

Article~17 of the AI Act requires providers of high-risk AI systems to establish, implement, document, and maintain a quality management system (QMS). The QMS ensures that the organisation applies consistent procedures for design, testing, data management, and post-market monitoring. It provides a structured approach for achieving and maintaining conformity with the requirements of the AI Act. Quality management thereby supports traceability, continuous improvement, and long-term compliance across the AI system lifecycle.

{\footnotesize
\begin{longtable}{|p{2.1cm}|p{2cm}|p{8cm}|p{0.8cm}|p{0.8cm}|}
\hline
\textbf{Requirement} & \textbf{References} & \textbf{Methods} & \textbf{Type} & \textbf{Target} \\
\hline
T8.1: Quality policy \& objectives & ISO~9001:2015 & M8.1: Quality Policy Declaration - Verify that top management has issued a formal, documented policy defining quality objectives and regulatory compliance commitments for high-risk AI systems under a Quality Management System (QMS). The QMS would determine, implement, and control processes needed for quality management, and maintain documented information to support operation and provide evidence of conformity.  & C  & P \\ 
\hline
T8.2: Documented QMS procedures & ISO~9001:2015 & M8.2: Control of documented information - Check that all lifecycle stages (design, development, validation, post-market) have controlled and versioned documented information stored in a document management system.
 & C & P\\ 
\hline
T8.3: Internal quality audits & ISO~9001:2015 & M8.3: Internal Auditing - Plan and perform periodic audits to verify conformity of processes and compliance with the Regulation; record results and corrective actions.  & C  & P \\ 
\hline
T8.4: Continuous improvement & ISO~9001:2015 & M8.4: Continuous Improvement - Use audit results, performance data, and feedback to identify non-conformities and improvement opportunities; update processes and objectives accordingly. & C & P \\
\hline
\end{longtable}
\vspace{-3mm}
}

\subsection{R9: Risk Management}

Article~9 of the AI Act establishes the obligation to implement a continuous and systematic risk management process throughout the lifecycle of high-risk AI systems. This process covers the identification, analysis, evaluation, and control of risks, as well as post-market monitoring and review. It must ensure that residual risks are acceptable relative to the intended use and that appropriate mitigation measures are applied. The methodology used here aligns with a set of horizontal, cross-sectoral standards, including ISO/IEC 23894 (AI risk management), ISO/IEC 42001 (AI management system), ISO 31000 and ISO 31010 (General risk management principles and techniques), ISO/IEC 22989 (AI concepts and terminology) as well as data- and robustness-related standards such as ISO/IEC 5259 series, ISO 8000 series, ISO/IEC TR 24027, ISO/IEC 24029-1/-2, and ISO/IEC 24970. In the future, when harmonised standards under the AI Act, particularly prEN 18228 (AI Risk Management) and prEN 18286 (AI Quality Management System) will be finalised, they will provide the solid structure for conformity verification and presumption of conformity once a system is categorised as high-risk.

However, it is essential to underline that none of these standards, both international ISO/IEC standards and future harmonised European standards, determine whether an AI system is classified as high-risk under Article 6. Classification is governed exclusively by the legal criteria of the AI Act, particularly in Annex III. Standards are used in this report solely to support structured analysis, documentation and evidence generation, and they become fully applicable only after an AI system has been classified as high-risk.
In addition, the methodology deliberately excludes sector-specific or application-specific standards, such as functional-safety standards (e.g., IEC 61508 or ISO 13849), medical-device standards (e.g., ISO 14971) or automotive safety standards (e.g., ISO 26262). These frameworks are highly relevant within their respective regulatory regimes but are not part of the horizontal, cross-sector risk-classification process defined by the AI Act and therefore are not included in the present analysis.

{\footnotesize
\begin{longtable}{|p{2.2cm}|p{4.6cm}|p{4.9cm}|p{0.8cm}|p{0.8cm}|}
\hline
\textbf{Requirement} & \textbf{References} & \textbf{Methods} & \textbf{Type} & \textbf{Target} \\
\hline
T9.1: System Definition &
ISO/IEC 42001 (\S\S~4.3, 6.1.2); ISO/IEC 23894 (\S~6.3); ISO/IEC 22989 &
M9.1 Description of AI system -- define intended purpose for Art 6 mapping and record system details &
C & P \\
\hline
T9.2: Risk identification &
ISO/IEC 42001 (\S\S~6.1.2, 7.5); ISO/IEC 23894 (\S\S~5.4, 6.4); ISO 31000; ISO 31010;  ISO/IEC 5259-1; ISO 8000-8; ISO/IEC TR 24027 &
M9.2 Structured risk identification relevant to classification (Art. 6) and risk management (Art. 9) &
C & P \\
\hline
T9.3: Risk analysis &
ISO/IEC 23894 (\S\S~6.4, 6.5); ISO/IEC 42001; ISO 31000; ISO 31010; ISO/IEC 24029-2 &
M9.3 Core classification step -- apply AI Act Art. 6: Annex I \& Annex III gates using structured analysis &
C & P \\
\hline
T9.4: Risk evaluation \& prioritisation &
ISO/IEC 23894 (\S\S~6.4, 6.5); ISO/IEC 42001; ISO 31000; ISO 31010 &
M9.4 Finalize classification decision;record justification and prioritise risks &
C & P \\
\hline
T9.5: Risk control considerations &
ISO/IEC 23894 (\S\S~5.5--5.7); ISO/IEC 42001; ISO 31000; ISO/IEC 24029-1/-2; ISO/IEC TR 24027; ISO/IEC 5259; ISO 8000; ISO/IEC 27001; ISO/IEC 27005 &
M9.5 Outline potential risk areas aligned with AI Act essential requirements &
C & FP \\
\hline
T9.6: PMM: Planning &
ISO/IEC 42001 (\S\S~8.4, 9.1, 7.5); ISO/IEC 23894 (\S\S~5.3, 5.7); ISO 31000;  ISO/IEC 24970; ISO/IEC 27001; ISO/IEC 27005 &
M9.6 Establishing PMM plan with monitoring, logging and reporting&
C & P \\
\hline
T9.7: PMM: Data collection \& analysis &
ISO/IEC 42001 (\S\S~9.1, 9.3, 7.5); ISO/IEC 23894 (\S\S~5.5--5.7); ISO 31000;  ISO/IEC 24970; ISO/IEC 27001; ISO/IEC 27005 &
M9.7 Continuous verification and early detection of anomalies or new hazards &
T & FP \\
\hline
T9.8: PMM: Review \& CAPA &
ISO/IEC 42001 (\S\S~8.3, 10, 7.5); ISO/IEC 23894 (\S\S~5.6, 5.8); ISO 31000 &
M9.8 Perform CAPA, update system, maintain continuous compliance  &
C & P \\
\hline
T9.9: Incident reporting &
ISO/IEC 42001 (\S\S~8.4, 9.1, 7.5); ISO/IEC 23894 (\S\S~5.7, 5.8);  ISO/IEC 24970; ISO/IEC 27035 &
M9.9 Compliant incident reporting; documentation; integrate lessons learned &
C & P \\
\hline
\end{longtable}
\vspace{-3mm}
}

\subsection{R10: Technical Documentation}

Article~11 and Annex~IV of the AI Act require providers of high-risk AI systems to prepare and maintain comprehensive technical documentation demonstrating compliance with the regulation. This documentation must provide sufficient detail to assess the system's conformity, including its intended purpose, design specifications, data sources, testing methods, and risk management processes. Proper documentation ensures transparency, enables traceability of design choices, and supports both internal governance and external conformity verification.

{\footnotesize
\begin{longtable}{|p{2.2cm}|p{4.6cm}|p{4.9cm}|p{0.8cm}|p{0.8cm}|}
\hline
\textbf{Requirement} & \textbf{References} & \textbf{Methods} & \textbf{Type} & \textbf{Target} \\
\hline
R10.1: System overview \& intended purpose & ISO/IEC 42001 (\S\S~4.3, 6.1.2, 7.5); ISO/IEC 23894 (\S~ 6.3); ISO/IEC 22989 & M10.1 System overview and intended purpose documented in alignment with Annex IV and QMS requirements. & C & FP\\ 
\hline
R10.2: Design specifications \& architecture & ISO/IEC 42001(\S\S~ 8.1--8.3, 7.5); ISO/IEC 23894 (\S\S~ 6.4--6.5)   & M10.2 Architecture and design specifications captured with complete traceability to risks and requirements. & C & D\&M\\ 
\hline
R10.3: Dataset description \& provenance &ISO/IEC 42001 (\S\S~ 7.5, 8.3); ISO/IEC 5259-1; ISO 8000-8; ISO/IEC TR 24027& M10.3 Dataset provenance, quality attributes and limitations documented in compliance with Art. 10--11 and Annex IV. & C & D\&M\\ 
\hline
R10.4: Performance metrics \& validation &ISO/IEC 42001 (\S\S~ 8.3, 9.1, 7.5); ISO/IEC 23894 (\S\S~~ 6.4–6.5); ISO/IEC 24029-1/-2; ISO/IEC 24970; ISO/IEC 27001; ISO/IEC 27005& M10.4 Performance, robustness and validation evidence compiled and traceable to requirements, risks and intended purpose. & T & FP\\ 
\hline
R10.5: Compliance evidence mapping &ISO/IEC 42001 (\S\S~ 6.1, 8.1--–8.4, 9.1–9.3, 10); ISO/IEC 23894 (\S\S~5.3--5.8); ISO/IEC 27001; ISO/IEC 27005& M10.5 Structured compliance mapping demonstrating how Annex IV and Article 17 obligations are satisfied. & C & P\\
\hline
\end{longtable}
\vspace{-3mm}
}

\subsection{R11: Record-keeping}

Article~18 of the AI Act mandates that providers retain automatically generated logs and relevant records to ensure traceability and accountability throughout the AI lifecycle. Record-keeping supports post-market monitoring, facilitates incident investigation, and provides evidence for conformity verifications. It must include documentation of model versions, dataset lineage, and data retention or deletion policies, consistent with organisational governance and data protection requirements, including GDPR.

{\footnotesize
\begin{longtable}{|p{2.2cm}|p{4.6cm}|p{4.9cm}|p{0.8cm}|p{0.8cm}|}
\hline
\textbf{Requirement} & \textbf{References} & \textbf{Methods} & \textbf{Type} & \textbf{Target} \\
\hline
R11.1: Operational logs &ISO/IEC 42001 (\S\S~ 7.5, 8.3, 9.1); ISO/IEC 24970 & M11.1 Operational logging system in place with records supporting Art. 18 record-keeping and PMM requirements. & C & FP \\ 
\hline
R11.2: Version control records &ISO/IEC 42001 (\S\S~ 7.5, 8.3) & M11.2 Version control ensures full traceability and reproducibility of AI system states over time. & C & D\&M \\ 
\hline
R11.3: Model \& dataset lineage &ISO/IEC 42001 (\S\S~ 7.5, 8.3); ISO/IEC 23894 (\S\S~5.3–5.8) & M11.3 Full lineage enables auditing, incident analysis, risk traceability and evidence-based updates. & C & D\&M\\ 
\hline
R11.4: Retention \& deletion policies &ISO/IEC 42001 (\S\S~ 7.5, 8.3); ISO/IEC 23894 (\S\S~5.3–5.8); ISO/IEC 27001; ISO/IEC 27005; GDPR (Art.~5.1.c, e)  & M11.4 Retention/deletion policies implemented to ensure compliance, minimise risk and maintain only necessary records, in line with GDPR data minimisation and storage limitation principles & C & P \\
\hline
\end{longtable}
\vspace{-3mm}
}

\section{Case Study}
\label{sec:case_study}

This section illustrates how the mapping proposed in Section~\ref{sec:mapping} can be applied in practice through a real, ongoing high-risk AI use case currently being analysed and tested in collaboration with Scania, a Swedish automotive manufacturer. While the mapping is generic and technology-agnostic, this example shows how it can structure assurance activities across the AI lifecycle, from design-time verification to validation and documentation. Table~\ref{tab:use_case} presents the results of this mapping process.
The use case concerns an AI-based system for detecting cyberattacks within connected vehicles in an industrial setting at Scania. Modern vehicles rely on internal communication networks that enable real-time data exchange between electronic components. Protecting these networks is critical, as cyberattacks may compromise both vehicle safety and system reliability. To address these risks, automotive manufacturers increasingly deploy Intrusion Detection Systems (IDS) that monitor in-vehicle communications and identify abnormal or potentially harmful behaviour.
In this setting, the IDS employs AI techniques to analyse in-vehicle network traffic and detect patterns associated with known or emerging attacks. The system processes network data, extracts relevant features, and uses machine-learning models to distinguish normal operation from suspicious activity. To support transparency and engineering oversight, explainability mechanisms are integrated to allow engineers to understand why specific activities are flagged as intrusions.
Designed for real-time deployment in vehicle control units or central gateways, this use case demonstrates the application of the proposed mapping in a complex, safety-critical automotive context. The mapping enables structured and traceable verification of data handling, model behaviour, explainability, and documentation, providing evidence to support safety, cybersecurity, and regulatory requirements without intervening in system development.

{\footnotesize
\begin{longtable}{|p{2cm}|p{2cm}|p{2cm}|p{8cm}|}
\caption{Results of the mapping instantiation for the Scania use case.}
\label{tab:use_case}\\
\hline
\textbf{Req.} & \textbf{Sub-req.} & \textbf{Method} & \textbf{Instantiation in the Use Case} \\
\hline
\endfirsthead
\hline
\textbf{Legal Requirement (AI Act)} & \textbf{Test / Control Type} & \textbf{Method} & \textbf{Instantiation in the Use Case} \\
\hline
\endhead

\multirow{6}{2cm}{R1: Human Agency \\ \& Oversight (Art. 14)}
& \multirow{3}{2cm}{T1.1 Human-in-the-loop}
& M1.1 Manual override
& Engineers and operators can override IDS decisions and reclassify labels during development.
\\ \cline{3-4}

&
& M1.2 Supervisory control
& IDS outputs are logged and verified by safety and security engineers before integration into final decisions.
\\ \cline{3-4}

&
& M1.3 Strategic governance
& Ensure a documented organisational oversight structure: define who in the team (e.g., vehicle OEM cybersecurity lead) is responsible for authorising deployment and change management of the IDS.
\\ \cline{2-4}

&
\multirow{2}{2cm}{T1.2 User Autonomy}
& M1.4 Informed consent
& Data collection from vehicles was conducted under user consent for research and model development.
\\ \cline{3-4}

&
& M1.5 User preferences
& Provide vehicle end-users or clients (e.g., fleet operators) the ability to configure IDS alert sensitivity or disable non-critical automated responses (while still keeping core safety monitoring active).
\\ \cline{2-4}

&
\multirow{1}{2cm}{T1.3 Oversight Mechanisms}
& M1.7 Audit trails
& The IDS decisions, rule activations, and system overrides are logged for auditability.
\\ \hline

\multirow{6}{2cm}{R2: Technical Robustness \\ \& Safety (Art. 15)}
& \multirow{3}{2cm}{T2.1 Resilience \& Reliability}
& M2.1 Stress testing
& Baseline models were evaluated under high-throughput simulated traffic conditions.
\\ \cline{3-4}

&
& M2.2 Drift detection
& Monitor for changes over time in the in-vehicle network traffic (e.g., ECU updates, different firmware) and detect when feature distributions deviate; trigger retraining or model review.
\\ \cline{3-4}

&
& M2.4 Fail-over architecture
& Ensure the IDS has a fallback mode: if the neuro-symbolic model fails or confidence drops, revert to a simpler rule-based guard or human-supervised monitoring.
\\ \cline{2-4}

&
\multirow{1}{2cm}{T2.2 Robustness to Attacks}
& M2.5 Red teaming
& The system was tested with synthetic adversarial attacks, including crafted PTP manipulations.
\\ \cline{2-4}

&
\multirow{1}{2cm}{T2.3 Fallback \& Fail-safe}
& M2.6 Human fallback
& Low-confidence detections trigger alerts requiring human verification.
\\ \cline{2-4}

&
\multirow{1}{2cm}{T2.4 Accuracy \& Uncertainty}
& M2.7 Model calibration
& Performance metrics include confidence intervals; rule activation thresholds are tunable.
\\ \hline

\multirow{3}{2cm}{R3: Privacy \& Data Governance (Art. 10; GDPR)}
& \multirow{1}{2cm}{T3.1 Data Minimisation}
& M3.1 Purpose limitation
& Only relevant features for detection (e.g., timestamp, ID, payload length) are processed.
\\ \cline{2-4}

&
\multirow{1}{2cm}{T3.2 Privacy-Enhancing}
& M3.3 Pseudonymisation
& When capturing vehicle network traffic for training, any identifiable driver or vehicle identity data is pseudonymised or removed, retaining only features necessary for intrusion detection.
\\ \cline{2-4}

&
\multirow{1}{2cm}{T3.3 Access \& Security}
& M3.4 Role-based access control
& Raw data is accessible only to authorised research engineers; results are pseudonymised.
\\ \hline

\multirow{5}{2cm}{R4: Transparency (Art. 13, 50)}
& \multirow{1}{2cm}{T4.1 Explainability}
& M4.1 SHAP / M4.2 LIME
& SHAP is used to validate feature contributions; LNN rules provide symbolic reasoning.
\\ \cline{2-4}

&
\multirow{2}{2cm}{T4.2 Model \& Data Traceability}
& M4.4 Model cards
& All models used are accompanied by documentation on training data, assumptions, and risks.
\\ \cline{3-4}

&
& M4.5 Datasheets
& Provide a datasheet for the dataset used in the IDS: describe origin (e.g., vehicles from the partner), format, attack types covered, limitations, and preprocessing steps to support traceability in audits.
\\ \cline{2-4}

&
\multirow{1}{2cm}{T4.3 Disclosure of AI Use}
& M4.6 AI identity notice
& In future deployments, HMI interfaces will indicate when AI-based IDS is active.
\\ \hline

\multirow{2}{2cm}{R5: Fairness \& Non-Discrimination (Art. 9, 10)}
& \multirow{1}{2cm}{T5.1 Fairness in Data}
& M5.1 Data bias detection
& Attack data were balanced to ensure representative detection performance.
\\ \cline{2-4}

&
\multirow{1}{2cm}{T5.4 Stakeholder Engagement}
& M5.5 Participatory design review
& Security experts and domain engineers reviewed and refined rule design and evaluation criteria.
\\ \hline

\multirow{5}{2cm}{R7: Accountability (Art. 12, 17)}
& \multirow{1}{2cm}{T7.1 Responsibility Assignment}
& M7.1 Responsibility-matrix verification
& Roles for data handling, model design, and validation are clearly documented.
\\ \cline{2-4}

&
\multirow{1}{2cm}{T7.2 Auditability}
& M7.2 Logging
& All detection events and rule-based decisions are logged and version controlled.
\\ \cline{2-4}

&
\multirow{1}{2cm}{T7.3 Incident Reporting}
& M7.4 Redress systems
& False positives or missed detections can be flagged through internal validation interfaces.
\\ \cline{2-4}

&
\multirow{1}{2cm}{T7.4 Documentation Integrity}
& M7.5 Version-controlled records
& Maintain version control for IDS models, rule sets, training datasets, and change logs (who changed what, when, and why) to support traceability and post-incident investigation.
\\ \hline

\multirow{4}{2cm}{R8: Quality Management (Art. 17)}
& \multirow{1}{2cm}{T8.1 Quality policy \& objectives}
& M8.1 Quality Policy Declaration
& The company’s cybersecurity and safety management define a documented QMS policy for AI-based IDS development, specifying objectives on detection accuracy, explainability, robustness, and regulatory alignment; relevant engineers receive QMS training.
\\ \cline{2-4}

&
\multirow{1}{2cm}{T8.2 Documented QMS procedures}
& M8.2 Control of documented information
& IDS development artifacts (model architectures, LNN rule sets, preprocessing pipelines, datasets, validation reports) are placed under controlled documentation with restricted access and traceability.
\\ \cline{2-4}

&
\multirow{1}{2cm}{T8.3 Internal quality audits}
& M8.3 Internal auditing
& Periodic internal audits review model lineage logs, robustness testing outcomes, fairness checks, and documentation completeness; non-conformities trigger corrective actions.
\\ \cline{2-4}

&
\multirow{1}{2cm}{T8.4 Continuous improvement}
& M8.4 Continuous improvement
& Findings from audits and monitoring are used to update models, rules, thresholds, and documentation throughout the lifecycle.
\\ \hline

\multirow{5}{2cm}{R9: Risk \\ classification, \\ management, \\ and post marketing  (Art. 6, Art. 9)}
& \multirow{1}{2cm}{T9.1 System Definition}
& M9.1 Description of AI system
& The automotive AI‑based IDS is defined as a system intended to detect cyber intrusions targeting in‑vehicle networks and connected services.
\\ \cline{2-4}

&
\multirow{1}{2cm}{T9.2 Risk identification}
& M9.2 Structured risk identification 
& Structured risk identification places the automotive IDS as preliminarily high‑risk due to safety‑critical vehicle functions.
\\ \cline{2-4}

&
\multirow{1}{2cm}{T9.3 Risk analysis}
& M9.3 Core classification step 
& The IDS is assessed against AIA Annex I \& III, with critical‑infrastructure and safety‑component gating confirming high‑risk applicability.
\\ \cline{2-4}

&
\multirow{1}{2cm}{T9.5 Risk control measures}
& M9.5 Risk control strategy
& Risk controls are established, covering safety, robustness, explainability, human oversight, and documentation through git platform in automotive contexts.
\\ \cline{2-4}

&
\multirow{1}{2cm}{T9.6 PMM planning}
& M9.6 PMM system established
& A post‑market monitoring plan is defined to track performance, drift, and safety‑relevant anomalies across vehicle fleets.
\\ \hline

\multirow{5}{2cm}{R10: Technical \\ Documentation \\ (Art. 11; Annex IV)}
& \multirow{1}{2cm}{T10.1 System overview \& intended purpose}
& M10.1 Annex-IV compliant description
& Provide a structured description of the IDS: architecture, model pipeline, LNN rule types, deployment targets (ECUs, gateways), and intended function (early anomaly detection).
\\ \cline{2-4}

&
\multirow{1}{2cm}{T10.2 Design specifications \& architecture}
& M10.2 Architecture traceability
& Document the full dataflow: CAN/Ethernet packet capture $\rightarrow$ preprocessing $\rightarrow$ feature extraction $\rightarrow$ ML/LNN inference $\rightarrow$ explainability layer $\rightarrow$ alert routing, with traceability from risks to design decisions.
\\ \cline{2-4}

&
\multirow{1}{2cm}{T10.3 Dataset description \& provenance}
& M10.3 Dataset provenance
& Record dataset sources (vehicle captures), attack types, collection conditions, preprocessing steps, firmware versions when relevant, and known limitations (e.g., limited coverage of rare timing attacks).
\\ \cline{2-4}

&
\multirow{1}{2cm}{T10.4 Performance metrics \& validation}
& M10.4 Validation evidence
& Provide detection rate, latency under high traffic, robustness under noise, calibration curves, and interpretability validation using SHAP and LNN reasoning.
\\ \cline{2-4}

&
\multirow{1}{2cm}{T10.5 Compliance evidence mapping}
& M10.5 Mapping to obligations
& Map each AI Act requirement to evidence: oversight logs, stress-test results, explainability artefacts, model cards, dataset datasheets, QMS records, and PMM outputs.
\\ \hline

\multirow{4}{2cm}{R11: Record-keeping \\ (Art. 12, 18)}
& \multirow{1}{2cm}{T11.1 Operational logs}
& M11.1 Operational logging
& Maintain logs of detections, LNN rule activations, SHAP explanations shown for review, fallback events, and red-team outcomes; store logs securely in automotive backend systems.
\\ \cline{2-4}

&
\multirow{1}{2cm}{T11.2 Version control records}
& M11.2 Version control
& Track versions of ML models, LNN logic rules, preprocessing code, and dataset snapshots in a versioned repository with approvals and traceable change history.
\\ \cline{2-4}

&
\multirow{1}{2cm}{T11.3 Model \& dataset lineage}
& M11.3 Lineage documentation
& Maintain lineage links from raw CAN/Ethernet captures to preprocessing scripts, training datasets, trained model versions, and deployed artefacts to support reproducibility and audit.
\\ \cline{2-4}

&
\multirow{1}{2cm}{T11.4 Retention \& deletion policies}
& M11.4 Retention/deletion
& Define retention periods for logs, dataset snapshots, and models; ensure GDPR compliance when identifiable metadata appears; enforce secure deletion mechanisms.
\\ \hline
\end{longtable}
}

\section{Discussion}
\label{sec:discussion}

Rather than functioning solely as an analytical construct, the proposed mapping is intended to support the institutional ecosystem through which AI Act compliance is operationalised in practice. By linking legal requirements to identifiable and verifiable assessment activities, the mapping provides a common reference structure that can be reused across provider self-assessment, regulatory sandbox experimentation, and formal conformity assessment. This addresses a central implementation risk of the AI Act: that identical legal obligations may be translated into materially different verification practices across organisations, sectors, and Member States.

For providers, the mapping enables structured self-assessment as part of continuous monitoring and post-market obligations. By making explicit which verification activities correspond to which legal requirements, it supports systematic internal evaluations over time, including reassessment following model updates, data drift, or changes in deployment context. This reduces reliance on implicit or ad hoc interpretations of compliance and facilitates the generation of traceable evidence that can be reused across internal governance, regulatory reporting, and external review.
In regulatory sandboxes established under Article~57, the mapping can function as a shared interpretive layer between developers and competent authorities. By clarifying how experimental testing activities relate to specific AI Act obligations, it supports the structuring of testing plans, documentation practices, and exit reports, while preserving the flexibility required for innovation-oriented assessment. This role is particularly relevant in a cross-border context, where divergent sandbox practices risk emerging in the absence of a common operational reference.
European-funded initiatives such as AI Factories \cite{eu_ai_factories} and Testing and Experimentation Facilities (TEFs) \cite{eu_tefs} are expected to play an increasingly central role in supporting the testing and validation of AI systems, including in the context of regulatory sandboxes, as reflected in recent policy and implementation guidance \cite{implementingAct}. In this setting, the availability of a shared mapping between legal requirements and verification activities becomes critical. These infrastructures are not merely technical providers, but intermediaries that translate regulatory expectations into concrete testing capabilities. A common mapping helps ensure that testing performed within such initiatives is interpretable, comparable, and reusable across regulatory and institutional contexts, rather than being tied to local or project-specific interpretations.

More broadly, the mapping contributes to convergence in verification methodologies across the EU by offering a reusable, technology-agnostic operational vocabulary. While enforcement remains decentralised and sectoral specificities must be respected, convergence on underlying verification logic is essential to avoid fragmentation of compliance practices. The mapping is therefore intentionally designed to be extensible: it can be refined, extended, and instantiated by different actors, including competent authorities, notified bodies, EU-funded infrastructures, and sectoral initiatives. Such extensions can introduce domain-specific metrics or procedures while preserving traceability to the original legal requirements.
In this sense, the mapping is best understood as an intermediate coordination layer between legal text and assessment practice. It does not prescribe how verification must be performed, but provides a common structure within which diverse practices can evolve in a coherent and interoperable manner. Encouraging its extension and sectoral instantiation is therefore not a limitation, but a necessary condition for achieving both regulatory consistency and practical relevance under the AI Act.

Finally, while this work is grounded in the European regulatory context and explicitly targets the implementation of the AI Act, the underlying challenge it addresses is not unique to Europe. Many jurisdictions are currently developing or refining AI-specific regulatory frameworks that similarly rely on high-level, principle-driven obligations whose practical verification remains underspecified. In this respect, the approach adopted in this paper is intended to be transferable. The notion of an explicit mapping between legal requirements and verifiable assessment activities can be adapted to other regulatory settings, supporting regulatory learning, comparability, and institutional capacity-building beyond the EU. The authors therefore encourage the development of analogous mappings in other jurisdictions and see value in future cross-jurisdictional dialogue on verification methodologies as AI regulation continues to evolve globally.

\section{Conclusion}
\label{sec:conclusion}

This paper introduced an explicit and structured mapping between high-level obligations under the EU AI Act and concrete, verifiable assessment activities. Starting from 11 high-level requirements used as a normative basis, the mapping decomposes these obligations into 48 operational sub-requirements and associates them with 66 verification activities. By making these relationships explicit, the mapping bridges the gap between regulatory intent and assessable technical and organisational evidence across the AI lifecycle.

A central contribution of this work is not merely the enumeration of verification activities, but the articulation of a reusable mapping logic. By organising verification activities along two orthogonal dimensions, i.e. the type of verification performed and the lifecycle target to which it applies, the mapping makes visible how compliance with the AI Act is constructed in practice. It shows that conformity emerges from structured combinations of procedural controls and empirical testing, rather than from isolated checks or single metrics.
By formalising these relationships, the mapping directly reduces key sources of uncertainty in AI Act implementation. Interpretive uncertainty is addressed by clarifying how abstract legal obligations can be decomposed into operational elements grounded in recognised practices. Operational uncertainty is reduced by identifying verification activities that are implementable, repeatable, and auditable across different organisational and technical contexts. Procedural uncertainty is mitigated by enabling consistent documentation, traceability, and comparison of evidence across systems, actors, and regulatory settings.

The value of the mapping extends beyond individual assessments. It provides a common reference that can be reused across provider self-assessment, regulatory sandboxes, conformity assessment, and post-market monitoring, supporting convergence in verification practices across Member States while preserving contextual flexibility. In doing so, it aligns with the AI Act’s governance model, which relies on continuous oversight and regulatory learning rather than one-off certification.
The mapping is intentionally technology-agnostic and non-prescriptive. Its purpose is not to fix verification practices in advance, but to offer a stable coordination layer between legal text and assessment practice that can be extended, refined, and instantiated by different actors, including sectoral initiatives and EU-funded testing infrastructures. In this sense, the contribution of this work lies in establishing a shared operational grammar for AI Act compliance verification, one that enables consistency without rigidity and innovation without fragmentation.

\section*{Acknowledgement}
This work is supported by the EU project Citcom.AI and Vinnova INTERSTICE project (reference number: 2024-00661). This work is also supported by the Swedish AI Factory and the Luxembourg AI Factory. 

\bibliographystyle{ACM-Reference-Format}
\bibliography{references}

@article{Brundage2020,
  author  = {Miles Brundage and et al.},
  title   = {Toward Trustworthy {AI} Development: Mechanisms for Supporting Verifiable Claims},
  journal = {arXiv preprint arXiv:2004.07213},
  year    = {2020}
}

@article{Basiri2016,
  author  = {Ali Basiri and Casey Rosenthal and Nora Jones and Andrew Hodges and Cole Mickens},
  title   = {Chaos Engineering},
  journal = {IEEE Software},
  volume  = {33},
  number  = {3},
  pages   = {35--41},
  year    = {2016}
}

@misc{mundhenk2020,
      title={Efficient Saliency Maps for Explainable AI}, 
      author={T. Nathan Mundhenk and Barry Y. Chen and Gerald Friedland},
      year={2020},
      eprint={1911.11293},
      archivePrefix={arXiv},
      primaryClass={cs.CV},
      url={https://arxiv.org/abs/1911.11293}, 
}

@article{Floridi2018,
  author  = {Luciano Floridi and Josh Cowls and et al.},
  title   = {{AI4People}: An Ethical Framework for a Good {AI} Society},
  journal = {Minds and Machines},
  volume  = {28},
  number  = {4},
  pages   = {689--707},
  year    = {2018}
}

@misc{GDPR2016,
  author       = {{European Union}},
  title        = {Regulation (EU) 2016/679 (General Data Protection Regulation)},
  howpublished = {Official Journal of the EU, L119},
  year         = {2016}
}

@misc{eu_ai_act_2024,
  title        = {Regulation (EU) 2024/1689 of the European Parliament and of the Council of 13 June 2024 laying down harmonised rules on artificial intelligence (Artificial Intelligence Act) and amending certain Union legislative acts},
  howpublished = {\url{https://eur-lex.europa.eu/legal-content/EN/TXT/?uri=CELEX:32024R1689}},
  note         = {OJ L 2024/1689, 12 July 2024},
  year         = {2024},
  month        = {July},
  institution  = {European Union},
}

@misc{eu_trustworthy_ai_2019,
  author       = {{High-Level Expert Group on Artificial Intelligence}},
  title        = {Ethics Guidelines for Trustworthy AI},
  year         = {2019},
  url          = {https://digital-strategy.ec.europa.eu/en/library/ethics-guidelines-trustworthy-ai},
  note         = {Accessed: 2025-05-25}
}

@standard{iso_iec_42001_2023,
  author       = {{International Organization for Standardization and International Electrotechnical Commission}},
  title        = {{ISO/IEC 42001:2023 – Artificial intelligence — Management system}},
  year         = {2023},
  howpublished = {\url{https://www.iso.org/standard/81230.html}},
  note         = {First AI Management System Standard, supporting transparency, fairness, and accountability},
}

@techreport{gpai-code-practice-3rd,
  title        = {General‑Purpose {AI} Code of Practice, Third Draft},
  author       = {{Chairs and Vice‑Chairs of the General‑Purpose AI Code of Practice}},
  institution  = {European AI Office / European Commission},
  type         = {Technical Draft},
  number       = {Draft 3},
  address      = {Brussels, Belgium},
  year         = {2025},
  month        = mar,
  day          = {11},
  url          = {https://digital-strategy.ec.europa.eu/en/library/third-draft-general-purpose-ai-code-practice-published-written-independent-experts},
  note         = {\emph{Third draft published for consultation; feedback invited until 30 March 2025, final version expected in May 2025.}},
}

@misc{unesco2021ethics,
  title        = {Recommendation on the Ethics of Artificial Intelligence},
  author       = {{UNESCO}},
  year         = {2021},
  howpublished = {\url{https://unesdoc.unesco.org/ark:/48223/pf0000381137}},
  note         = {Adopted on 23 November 2021 by the General Conference of UNESCO at its 41st session},
  institution  = {United Nations Educational, Scientific and Cultural Organization}
}

@book{isaca2025aaia,
  author       = {{ISACA}},
  title        = {AAIA Official Review Manual},
  subtitle     = {Advanced in AI Audit Certification},
  year         = {2025},
  publisher    = {ISACA},
  address      = {Rolling Meadows, IL},
  month        = may,
  isbn         = {979-8892270472},
  note         = {Print version, 182 pages; first released May 19, 2025}  
}

@misc{mowla2024ai,
  title={From AI Act to Structured Testing of AI Systems},
  author={Nishat Mowla and Fahria Kabir},
  year={2024},
  publisher={RISE Research Institutes of Sweden}
}

@misc{CEN-CENELEC_JTC21_2025,
  author       = {{CEN-CENELEC Joint Technical Committee 21}},
  title        = {European AI Standardization | CEN-CENELEC JTC 21},
  year         = {2025},
  howpublished = {\url{https://jtc21.eu}},
}

@misc{NIST_AI_RMF_2025,
  author       = {{National Institute of Standards and Technology (NIST)}},
  title        = {AI Risk Management Framework (AI RMF)},
  year         = {2025},
  howpublished = {\url{https://www.nist.gov/itl/ai-risk-management-framework}},
}

@article{deloitte2024ai,
  title={EU AI Act Survey: Uncertainty in Implementation},
  author={Deloitte},
  year={2024},
  journal={Deloitte Legal Research},
  url={https://www.deloitte.com/dl/en/services/legal/research/umfrage-eu-ai-act-2024.html},
}

@article{dlapiper2025pause,
  title={The European Commission Considers Pause on AI Act's Entry into Application},
  author={DLA Piper},
  year={2025},
  journal={AI Outlook Report},
  url={https://www.dlapiper.com/en/insights/publications/ai-outlook/2025/the-european-commission-considers-pause-on-ai-act-entry-into-application},
}

@report{draghi2024competitiveness,
  title={EU Competitiveness Report (Draghi Report)},
  author={Draghi, Mario},
  year={2024},
  institution={European Commission},
  url={https://sciencebusiness.net/news/ai/eu-losing-narrative-battle-over-ai-act-says-un-adviser},
}

@article{arnal2024implementation,
  title={AI at Risk in the EU: It’s Not Regulation, It’s Implementation},
  author={Arnal, Julien},
  journal={European Journal of Risk Regulation},
  year={2024},
  publisher={Cambridge University Press},
  url={https://www.cambridge.org/core/journals/european-journal-of-risk-regulation/article/ai-at-risk-in-the-eu-its-not-regulation-its-implementation/A9FD120F3EACE2C083048ABCBF96C0F6},
}

@misc{lewis2025regulatorylearning,
  title={Mapping the Regulatory Learning Space for the EU AI Act},
  author={Lewis, Dave and Lasek-Markey, Maria and Golpayegani, Donya and Pandit, Harshvardhan J.},
  year={2025},
  howpublished={arXiv preprint arXiv:2503.05787},
  url={https://arxiv.org/abs/2503.05787},
}

@article{regulatinguncertainty2025,
  title={Regulating Uncertainty: Governing General-Purpose AI Models and Systemic Risk},
  author={unknown},
  year={2025},
  journal={European Journal of Risk Regulation},
  publisher={Cambridge University Press},
  url={https://resolve.cambridge.org/core/journals/european-journal-of-risk-regulation/article/regulating-uncertainty-governing-generalpurpose-ai-models-and-systemic-risk/7EEFE1D8421A43A98CE91F7C697DE538},
}

@standard{ISO23894,
  title        = {{Information technology -- Artificial intelligence -- Guidance on risk management}},
  organization = {International Organization for Standardization (ISO) and International Electrotechnical Commission (IEC)},
  year         = {2023},
  number       = {ISO/IEC 23894:2023},
  url          = {https://www.iso.org/standard/77304.html},
  note         = {Available at: https://www.iso.org/standard/77304.html}
}

@standard{ISO9001,
  title = {{ISO 9001:2015 - Quality management systems — Requirements}},
  year = {2015},
  institution = {International Organization for Standardization (ISO)},
  address = {Geneva, Switzerland},
  number = {9001},
  url = {https://www.iso.org/standard/62085.html},
  note = {Latest version published in 2015}
}

@article{bengio2024managing,
  title={Managing extreme AI risks amid rapid progress},
  author={Bengio, Yoshua and Hinton, Geoffrey and Yao, Andrew and Song, Dawn and Abbeel, Pieter and Darrell, Trevor and Harari, Yuval Noah and Zhang, Ya-Qin and Xue, Lan and Shalev-Shwartz, Shai and others},
  journal={Science},
  volume={384},
  number={6698},
  pages={842--845},
  year={2024},
  publisher={American Association for the Advancement of Science}
}

@article{wang2022artificial,
  title={Artificial intelligence in safety-critical systems: a systematic review},
  author={Wang, Yue and Chung, Sai Ho},
  journal={Industrial Management \& Data Systems},
  volume={122},
  number={2},
  pages={442--470},
  year={2022},
  publisher={Emerald Publishing Limited}
}

@article{huang2024,
  title={AI regulations},
  author={Huang, Ken and Joshi, Aditi and Dun, Sandy and Hamilton, Nick},
  booktitle={Generative AI security: theories and practices},
  pages={61--98},
  year={2024},
  publisher={Springer}
}

@article{smuha2021race,
  title={From a ‘race to AI’to a ‘race to AI regulation’: regulatory competition for artificial intelligence},
  author={Smuha, Nathalie A},
  journal={Law, Innovation and Technology},
  volume={13},
  number={1},
  pages={57--84},
  year={2021},
  publisher={Taylor \& Francis}
}

@article{novelli2024taking,
  title={Taking AI risks seriously: a new assessment model for the AI Act},
  author={Novelli, Claudio and Casolari, Federico and Rotolo, Antonino and Taddeo, Mariarosaria and Floridi, Luciano},
  journal={Ai \& Society},
  volume={39},
  number={5},
  pages={2493--2497},
  year={2024},
  publisher={Springer}
}

@article{holtzmanGenerativeModelsComplex2025,
  title = {Generative {{Models}} as a {{Complex Systems Science}}: {{How Can We Make Sense}} of {{Large Language Model Behavior}}?},
  shorttitle = {Generative {{Models}} as a {{Complex Systems Science}}},
  author = {Holtzman, Ari and West, Peter and Zettlemoyer, Luke},
  year = 2025,
  month = jun,
  journal = {Journal of Social Computing},
  volume = {6},
  number = {2},
  pages = {75--94},
  issn = {2688-5255},
  doi = {10.23919/JSC.2025.0009},
}

@article{koltLessonsComplexSystems2025a,
  title = {Lessons from Complex Systems Science for {{AI}} Governance},
  author = {Kolt, Noam and {Shur-Ofry}, Michal and Cohen, Reuven},
  year = 2025,
  month = aug,
  journal = {Patterns},
  volume = {6},
  number = {8},
  pages = {101341},
  issn = {2666-3899},
  doi = {10.1016/j.patter.2025.101341},
  langid = {english},
}

@article{zakiConceptualisingOrganisationalPolicy2025,
  title = {Conceptualising Organisational Policy Learning: {{Triggers}}, Processes, Outcomes, and Implications for Policy and Governance Change},
  shorttitle = {Conceptualising Organisational Policy Learning},
  author = {Zaki, Bishoy},
  year = 2025,
  month = nov,
  journal = {Australian Journal of Public Administration},
  doi = {10.1111/1467-8500.70031},
}

@misc{schrepelAdaptiveRegulation2025,
  type = {{{SSRN Scholarly Paper}}},
  title = {Adaptive {{Regulation}}},
  author = {Schrepel, Thibault},
  year = 2025,
  month = aug,
  number = {5416454},
  eprint = {5416454},
  publisher = {Social Science Research Network},
  address = {Rochester, NY},
  doi = {10.2139/ssrn.5416454},
  urldate = {2025-11-20},
  archiveprefix = {Social Science Research Network},
  langid = {english},
}

@standard{ISO_IEC_22989_2022,
  type        =       {Standard},
  title       =       {{Information technology — Artificial intelligence — Artificial intelligence concepts and terminology}},
  organization=       {International Organization for Standardization and International Electrotechnical Commission},
  institution =       {ISO/IEC},
  number      =       {22989:2022},
  year        =       {2022},
  address     =       {Geneva, Switzerland},
  note        =       {ISO/IEC 22989:2022},
}

@standard{ISO_IEC_TR_24027_2021,
  type        =       {Technical Report},
  title       =       {{Information technology — Artificial intelligence (AI) — Bias in AI systems and AI-aided decision making}},
  organization=       {International Organization for Standardization and International Electrotechnical Commission},
  institution =       {ISO/IEC},
  number      =       {TR 24027:2021},
  year        =       {2021},
  address     =       {Geneva, Switzerland},
  note        =       {ISO/IEC TR 24027:2021},
}

@standard{ISO_IEC_24029_2_2023,
  type        =       {Standard},
  title       =       {{Artificial intelligence (AI) — Assessment of the robustness of neural networks — Part 2: Methodology for the use of formal methods}},
  organization=       {International Organization for Standardization and International Electrotechnical Commission},
  institution =       {ISO/IEC},
  number      =       {24029-2:2023},
  year        =       {2023},
  address     =       {Geneva, Switzerland},
  note        =       {ISO/IEC 24029-2:2023},
}

@standard{ISO_IEC_24970_DIS_2025,
  type        =       {Draft Standard},
  title       =       {{Artificial intelligence — AI System Logging}},
  organization=       {International Organization for Standardization and International Electrotechnical Commission},
  institution =       {ISO/IEC},
  number      =       {DIS 24970:2025},
  year        =       {2025},
  address     =       {Geneva, Switzerland},
}

@standard{ISO_IEC_27001_2022,
  type        =       {Standard},
  title       =       {{Information security, cybersecurity and privacy protection — Information security management systems — Requirements}},
  organization=       {International Organization for Standardization and International Electrotechnical Commission},
  institution =       {ISO/IEC},
  number      =       {27001:2022},
  year        =       {2022},
  address     =       {Geneva, Switzerland},
  note        =       {ISO/IEC 27001:2022},
}

@standard{ISO_IEC_27005_2022,
  type        =       {Standard},
  title       =       {{Information security, cybersecurity and privacy protection — Information security risk management}},
  organization=       {International Organization for Standardization and International Electrotechnical Commission},
  institution =       {ISO/IEC},
  number      =       {27005:2022},
  year        =       {2022},
  address     =       {Geneva, Switzerland},
  note        =       {ISO/IEC 27005:2022},
}

@standard{ISO_IEC_27035_1_2023,
  type        =       {Standard},
  title       =       {{Information security, cybersecurity and privacy protection — Information security incident management — Part 1: Principles of incident management}},
  organization=       {International Organization for Standardization and International Electrotechnical Commission},
  institution =       {ISO/IEC},
  number      =       {27035-1:2023},
  year        =       {2023},
  address     =       {Geneva, Switzerland},
  note        =       {ISO/IEC 27035-1:2023},
}

@standard{ISO_31000_2018,
  type        =       {Standard},
  title       =       {{Risk management — Guidelines}},
  organization=       {International Organization for Standardization},
  institution =       {ISO},
  number      =       {31000:2018},
  year        =       {2018},
  address     =       {Geneva, Switzerland},
  note        =       {ISO 31000:2018},
}

@standard{ISO_IEC_31010_2019,
  type        =       {Standard},
  title       =       {{Risk management — Risk assessment techniques}},
  organization=       {International Electrotechnical Commission and International Organization for Standardization},
  institution =       {IEC/ISO},
  number      =       {31010:2019},
  year        =       {2019},
  address     =       {Geneva, Switzerland},
  note        =       {IEC 31010:2019},
}

@standard{ISO_IEC_5259_1_2024,
  type        =       {Standard},
  title       =       {{Artificial intelligence — Data quality for analytics and machine learning (ML) — Part 1: Overview, terminology and examples}},
  organization=       {International Organization for Standardization and International Electrotechnical Commission},
  institution =       {ISO/IEC},
  number      =       {5259-1:2024},
  year        =       {2024},
  address     =       {Geneva, Switzerland},
  note        =       {ISO/IEC 5259-1:2024},
}

@standard{ISO_8000_8_2015,
  type        =       {Standard},
  title       =       {{Data quality — Part 8: Information and data quality: Concepts and measuring}},
  organization=       {International Organization for Standardization},
  institution =       {ISO},
  number      =       {8000-8:2015},
  year        =       {2015},
  address     =       {Geneva, Switzerland},
  note        =       {ISO 8000-8:2015},
}

@article{hernandez2025open,
  title={An open knowledge graph-based approach for mapping concepts and requirements between the eu ai act and international standards},
  author={Hernandez, Julio and Golpayegani, Delaram and Lewis, Dave},
  journal={AI and Ethics},
  pages={1--12},
  year={2025},
  publisher={Springer}
}

@book{sheridan1992telerobotics,
  title={Telerobotics, automation, and human supervisory control},
  author={Sheridan, Thomas B},
  year={1992},
  publisher={MIT press}
}

@article{andreotta2022ai,
  title={AI, big data, and the future of consent},
  author={Andreotta, Adam J and Kirkham, Nin and Rizzi, Marco},
  journal={Ai \& Society},
  volume={37},
  number={4},
  pages={1715--1728},
  year={2022},
  publisher={Springer}
}

@techreport{wcag22,
  author       = {{World Wide Web Consortium (W3C)}},
  title        = {Web Content Accessibility Guidelines (WCAG) 2.2},
  institution  = {World Wide Web Consortium},
  year         = {2023},
  month        = oct,
  url          = {https://www.w3.org/TR/WCAG22/},
  note         = {W3C Recommendation}
}

@book{schuler1993participatory,
  title={Participatory design: Principles and practices},
  author={Schuler, Douglas and Namioka, Aki},
  year={1993},
  publisher={CRC press}
}

@misc{CoE_AI_2020,
  author={Leslie, David and Burr, Christopher and Aitken, Mhairi and Cowls, Josh and Katell, Michael and Briggs, Morgan},
  title  = {Human Rights, Democracy and the Rule of Law in the Age of Artificial Intelligence},
  year   = {2020},
  publisher={Council of Europe},
  url    = {https://search.coe.int/cm/Pages/result_details.aspx?ObjectID=09000016809c4bd1}
}

@article{de2022artificial,
  title={An artificial intelligence life cycle: From conception to production},
  author={De Silva, Daswin and Alahakoon, Damminda},
  journal={Patterns},
  volume={3},
  number={6},
  year={2022},
  publisher={Elsevier}
}

@article{wang2021artificial,
  title={Artificial intelligence in product lifecycle management},
  author={Wang, Lei and Liu, Zhengchao and Liu, Ang and Tao, Fei},
  journal={The International Journal of Advanced Manufacturing Technology},
  volume={114},
  number={3},
  pages={771--796},
  year={2021},
  publisher={Springer}
}

@misc{eu_tefs,
  title        = {Sectorial AI Testing and Experimentation Facilities under the Digital Europe Programme},
  howpublished = {\url{https://digital-strategy.ec.europa.eu/en/policies/testing-and-experimentation-facilities}},
  year         = {2025},
  institution  = {European Commission},
}

@misc{eu_ai_factories,
  title        = {{AI Factories - Shaping Europe's Digital Future}},
  howpublished = {\url{https://digital-strategy.ec.europa.eu/en/policies/ai-factories}},
  year         = {2025},
  author       = {{European Commission}},
}

@misc{implementingAct,
  author       = {{European Commission}},
  title        = {Draft - Implementing Act on AI regulatory sandboxes under the Artificial Intelligence Act},
  year         = {2025},
  howpublished = {\url{Draft Implementing Act AI regulatory sandboxes}},
}

\end{document}